\documentclass[aps,prl,floatfix,reprint,amsmath,amssymb,superscriptaddress]{revtex4-2}

\usepackage{amsfonts}
\usepackage{mathrsfs}
\usepackage{amsmath}
\usepackage{color}
\usepackage{natbib}
\usepackage{graphicx}
\usepackage{bm}
\usepackage{amssymb}
\usepackage{xspace}
\usepackage{epstopdf}
\usepackage{dcolumn}
\usepackage{longtable}
\usepackage{multirow}

\usepackage[colorlinks=true, letterpaper=true, pdfstartview=FitV, linkcolor=blue, citecolor=blue, urlcolor=blue]{hyperref}

\begin{document}

\preprint{APS/123-QED}

\title{Phonon-Dominated Thermal Transport and Large Violation of the Wiedemann-Franz Law in Topological Semimetal CoSi}
\author{Luyao Zhong}%
\affiliation{College of Physics, Chongqing University, Chongqing, China}
\affiliation{Center of Quantum Materials and Devices, Chongqing University, Chongqing, China}

\author{Xin Jin}
\affiliation{College of Physics and Electronic Engineering, Chongqing Normal University, Chongqing 401331, China}

\author{Mingquan He}
\affiliation{Low Temperature Physics Laboratory, College of Physics \& Center of Quantum Materials and Devices, Chongqing University, Chongqing 401331, China}

\author{Rui Wang}
\affiliation{College of Physics, Chongqing University, Chongqing, China}
\affiliation{Center of Quantum Materials and Devices, Chongqing University, Chongqing, China}

\author{Xiaoyuan Zhou}
\affiliation{College of Physics, Chongqing University, Chongqing, China}
\affiliation{Center of Quantum Materials and Devices, Chongqing University, Chongqing, China}

\author{Tianqi Deng}
\email{dengtq@zju.edu.cn}
\affiliation{State Key Laboratory of Silicon and Advanced Semiconductor Materials, School of Materials Science and Engineering, Zhejiang University, Hangzhou, China}
\affiliation{Key Laboratory of Power Semiconductor Materials and Devices of Zhejiang Province, Institute of Advanced Semiconductors, ZJU-Hangzhou Global Scientific and Technological Innovation Center, Hangzhou, China}

\author{Xiaolong Yang}%
 \email{yangxl@cqu.edu.cn}
\affiliation{College of Physics, Chongqing University, Chongqing, China}
\affiliation{Center of Quantum Materials and Devices, Chongqing University, Chongqing, China}

\date{\today}

\begin{abstract}
The Wiedemann-Franz (WF) law, relating the electronic thermal conductivity ($\kappa_{\rm e}$) to the electrical conductivity, is vital in numerous applications such as in the design of thermoelectric materials and in the experimental determination of the lattice thermal conductivity ($\kappa_{\rm L}$). While the WF law is generally robust, violations are frequently observed, typically manifesting in a reduced Lorenz number ($L$) relative to the Sommerfeld value ($L_0$) due to inelastic scattering. Here, we report a pronounced departure from the WF law in the topological semimetal CoSi, where the electronic Lorenz number ($L_{\rm e}$) instead rises up to \(\sim40\%\) above $L_0$. We demonstrate that this anomaly arises from strong bipolar diffusive transport, enabled by topological band-induced electron-hole compensation, which allows electrons and holes to flow cooperatively and additively enhance the heat current. Concurrently, we unveil that the lattice contribution to thermal conductivity is anomalously large and becomes the dominant component below room temperature. As a result, if $\kappa_{\rm L}$ is assumed negligible---as conventional in metals, the resulting $L$ from the total thermal conductivity ($\kappa_{\rm tot}=\kappa_{\rm L}+\kappa_{\rm e}$) deviates from $L_0$ by more than a factor of three. Our work provides deeper insight into the unconventional thermal transport physics in topological semimetals. 
\end{abstract}

\maketitle

\begin{figure*}
\centering  
\includegraphics[scale=0.6]{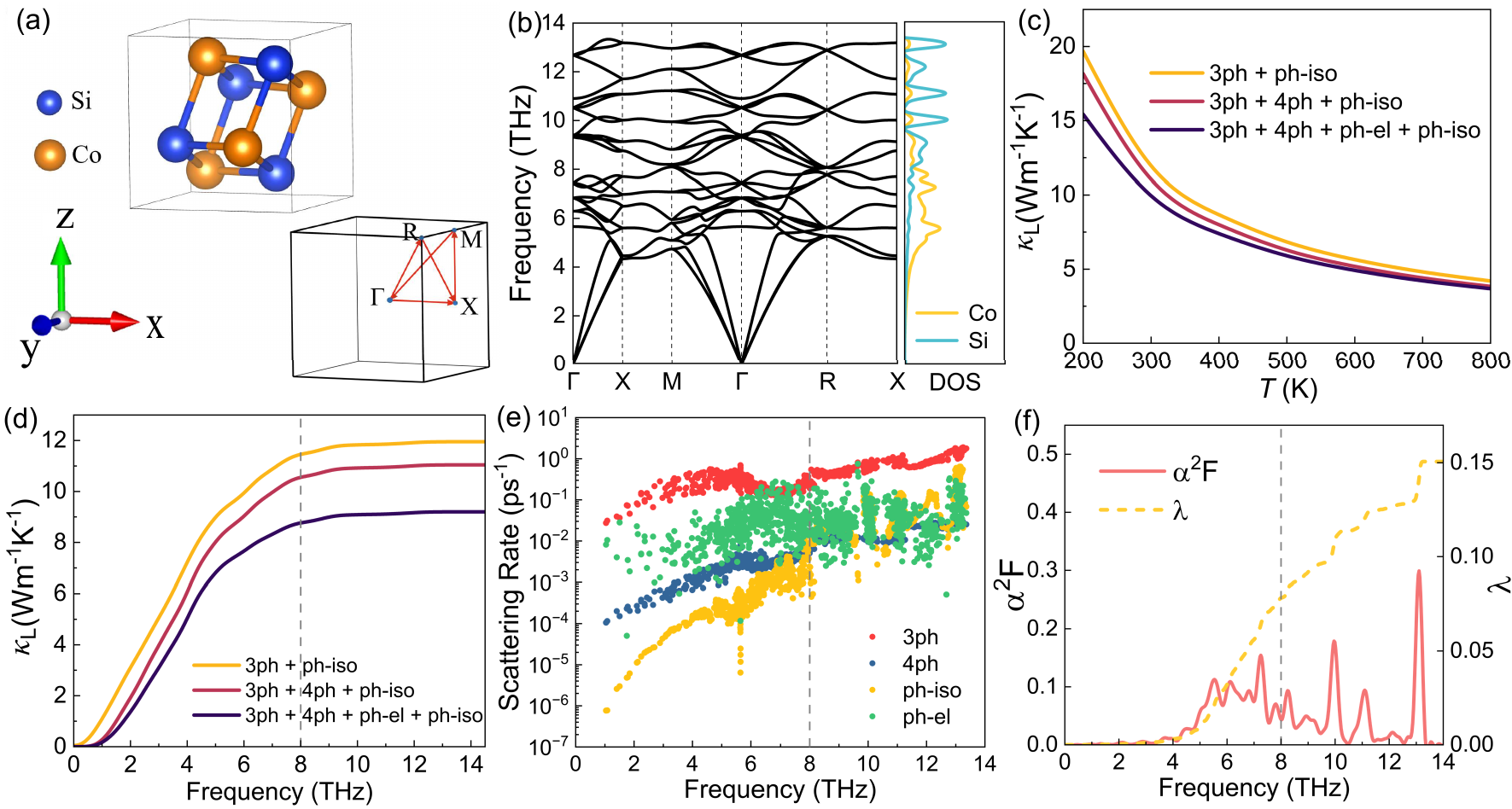}
\caption{(a) The crystal structure and the first Brillouin zone of CoSi, (b) the phonon dispersion along with the projected density of states, (c) the calculated $T$-dependent $\kappa_{\rm L}$, (d) cumulative lattice thermal conductivity considering different scattering mechanisms at 300 K, (e) the scattering rates contributed by the 3ph, 4ph, ph-iso, and ph-el processes, and (f) the Eliashberg spectral function $\alpha^2F(\omega)$ and el-ph coupling strength ($\lambda(\omega)$).}
\label{Fig1}
\end{figure*}

In metals, electrons serve as the dominant carriers for both charge and heat. Accordingly, the electronic thermal conductivity ($\kappa_{\rm e}$) and electrical conductivity ($\sigma$) are related through the Wiedemann-Franz (WF) law \cite{Franz1853, ashcroft1976solid, 5392734}: $\kappa_{\rm e} = L\sigma T$, where $T$ is the absolute temperature and $L$ is the Lorenz number. Knowledge of $L$ is not only of fundamental theoretical importance but also critical for numerous applications, including the design of thermoelectric materials and the experimental extraction of the lattice thermal conductivity ($\kappa_{\rm L}$). However, experimentally determination of $L$ usually poses challenges due to the difficulty in disentangling the electronic and lattice (phonon) contributions to the total thermal conductivity ($\kappa_{\rm tot}$). On the theoretical side, recent advancements in density functional theory (DFT) combined with the Boltzmann transport equation (BTE) have enabled accurate first-principles predictions of both $\kappa_{\rm e}$ and $\kappa_{\rm L}$ \cite{LI20141747, han2022fourphonon, li2015electrical, PhysRevB.99.020305, yang2021tuning, yang2021indirect, ding2024anharmonicity, ding2025concurrent}, thereby providing a powerful alternative route to determining $L$.

For most conventional metals and heavily doped semiconductors, where charge carriers behave as nearly free electrons, $L$ is well approximated by the Sommerfeld value $L_0=(\pi^2/3)(k_{\rm B}/e)^2= 2.44\times 10^{-8}$ W$\Omega\mathrm{K^{-2}}$ as derived from Fermi liquid theory \cite{Franz1853,ziman2001electrons,tulipman2023criterion, landau1957theory}. Here, $k_{\rm B}$ is the Boltzmann constant and $e$ denotes the elementary charge. Nevertheless, violations of the WF law have been widely identified both theoretically and experimentally \cite{kane1996thermal, garg2009large, kim2009violation, science.aad0343,PhysRevB.105.115113,PhysRevB.80.104510}, with the magnitude and even sign of deviation exhibiting pronounced material dependency. Such violations manifest as values of $L$ significantly below $L_0$ in strongly correlated electron systems \cite{dong2013anomalous,pfau2012thermal,science.ade3232}, where the absence of quasiparticles leads to independent diffusion of heat and charge, as well as upward deviations in metals where the thermal conductivity of the lattice is relatively large \cite{KUNDU2020100214,PhysRevB.99.020305}. More importantly, upward deviations can also arise from a fundamentally different origin: a dramatically enhanced electronic Lorenz number ($L_e$) due solely to novel electron transport physics, which have been attributed to various mechanisms such as non-Fermi liquid behavior and collective electron hydrodynamics in the low-temperature regime \cite{jaoui2018departure,Wang2025,PhysRevB.78.085416,PhysRevB.88.245444,PhysRevB.91.035414,PhysRevLett.115.056603,PhysRevB.92.115426,science.aad0343,PhysRevB.97.045105,PhysRevB.97.121405,PhysRevB.97.121404,PhysRevB.97.245128,PhysRevB.99.161407,PhysRevB.101.045421,PhysRevResearch.2.023391,Narozhny2022} and bipolar diffusive transport in the intermediate-temperature regime \cite{10.1063/1.5009939, https://doi.org/10.1002/adma.201404738,PhysRevB.95.125206, 10.1063/1.1728587,10.1063/1.1729056,https://doi.org/10.1002/pssb.2220650237,10.1063/1.3493269,doi:10.7566/JPSJ.84.024601,adfm.201902437,chen2021leveraging}. These departures not only provide a window into unconventional transport physics but also suggest routes toward decoupling charge and heat transport, with important implications for applications such as thermoelectrics. In parallel, topological semimetals have attracted intense interest over the past decade as a host of novel quantum transport phenomena emerging from their nontrivial band structures \cite{hasan2010colloquium, hu2019transport, narang2021topology, sakai2020iron}. However, thermal transport in these systems remains poorly understood. In particular, it is largely unknown whether they host unconventional heat transport mechanisms in the intermediate-temperature regime and how these may be linked to their topological band natures.

In this Letter, we perform first-principles calculations based on the accurate solution of BTE to study electrical and phonon transport in CoSi, a topological semimetal that has garnered significant attention due to its multiple types of topological fermions and unique surface states \cite{PhysRevLett.119.206402,Pshenay-Severin_2018,PhysRevLett.122.076402,pnas.2010752117,PhysRevB.106.224304,PhysRevLett.131.116603,PhysRevB.107.125145,PhysRevB.109.205115,PhysRevLett.129.026401}. Our results reveal an anomalously large, and even dominant, lattice contribution to the thermal conductivity at intermediate and low temperatures. Furthermore, we observe a drastic breakdown of the WF law in CoSi, manifested by a strongly enhanced Lorenz number far exceeding $L_0$. We attribute this violation to electron-hole compensation arising from topological band structure, which leads to pronounced bipolar diffusive transport. Our work establishes bipolar transport via topological band engineering as a general mechanism for decoupling charge and heat transport in solids.

Our first-principles calculations are performed within the framework of DFT and density functional perturbation theory, as implemented in the QUANTUM ESPRESSO package \cite{QE-2017}. We employ optimized norm-conserving pseudopotentials under the generalized gradient approximation \cite{PhysRevB.88.085117}, and include spin-orbit coupling (SOC) in all calculations. The $\kappa_{\rm L}$ is calculated by exactly solving the phonon BTE using the modified FourPhonon code \cite{LI20141747,han2022fourphonon}, taking into account scattering contributions from isotopes, three-phonon (3ph), four-phonon (4ph), and phonon-electron (ph-el) processes. The ph-el scattering rates are computed using the EPW package \cite{Lee2023}. All electrical transport properties are computed with the modified Perturbo code \cite{ZHOU2021107970}. Further computational details are provided in the Supplemental Material \cite{suppl}.

Figure 1(a) depicts the crystal structure and first Brillouin zone of CoSi. CoSi crystallizes in the simple cubic B20 structure (space group $P2_13$, No. 198), characterized by tetrahedral coordination with each Si atom bonded to four Co atoms within the unit cell. The optimized lattice parameter is 4.429 {\AA}, in agreement with experimental \cite{PhysRevB.82.155124,PhysRevB.86.064433} and previous theoretical values \cite{Pshenay-Severin_2018,PhysRevB.106.224304}. We begin by examining the intrinsic phonon thermal transport. The calculated phonon dispersion in Fig.~\ref{Fig1}(b) shows sharply dispersive acoustic branches with high group velocities, predominantly contributed by Co atoms as indicated by the phonon density of states (PDOS), which facilitates a relatively high $\kappa_{\rm L}$. As shown in Fig.~\ref{Fig1}(c), the calculated $T$-dependent $\kappa_{\rm L}$ is only weakly affected by both 4ph and ph-el scattering. The spectral contribution to $\kappa_{\rm L}$ in Fig.~\ref{Fig1} (d) reveals that phonons below 8 THz, encompassing all acoustic modes and low-lying optical branches, provide the dominant contribution. In this frequency regime, Fig.~\ref{Fig1}(e) confirms that 3ph processes dominate the phonon scattering, while 4ph and ph-el scatterings are rather weak, consistent with the weak suppression of $\kappa_{\rm L}$ observed in Fig.~\ref{Fig1} (c). The weak ph-el scattering, which is a key factor contributing to the relatively high $\kappa_{\rm L}$, is further corroborated by the spectral function and el-ph coupling strength ($\lambda$) shown in Fig.~\ref{Fig1}(f). Notably, the cumulative $\lambda$ when truncated at 8 THz is $\sim$0.07, significantly lower than that of typical metals \cite{PhysRevB.100.144306,GIRI2020100175}.

\begin{figure}
\includegraphics[scale=0.4]{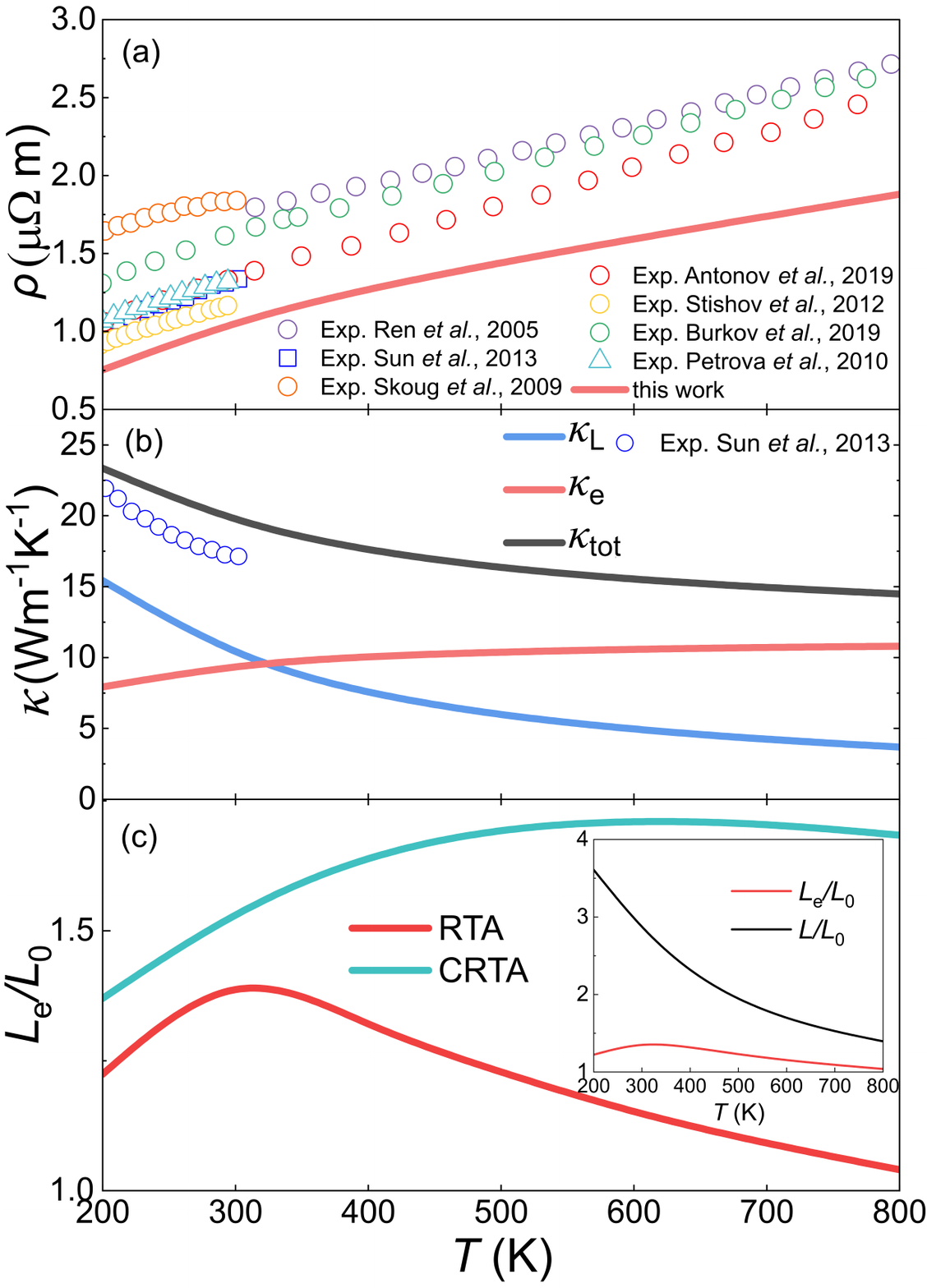}
\caption{(a) The electrical resistivity, (b) the thermal conductivity ($\kappa_{\rm e}$, $\kappa_{\rm L}$, and $\kappa_{\rm tot}$,), and (c) the Lorenz number with the inset showing the $L$ extracted from $\kappa_{\rm tot}$. The symbols show experimental electrical resistivity taken from Refs.\cite{REN200550, 10.1063/1.3072799, 10.1063/1.5119209, PhysRevB.86.064433, BURKOV2019540, PhysRevB.82.155124}, and thermal conductivity taken from Refs.\cite{Sun2013}.}
\label{Fig2}
\end{figure}

$\kappa_{\rm L}$ is inaccessible to direct experimental measurement, necessitating comparison between calculated and experimental total thermal conductivity ($\kappa_{\rm tot}=\kappa_{\rm e}+\kappa_{\rm L}$). To facilitate this comparison, we also calculate electronic transport properties including electrical resistivity $\rho$ and $\kappa_{\rm e}$. The calculated phonon-limited $\rho$ is displayed in Fig.~\ref{Fig2}(a), benchmarked against available experimental data \cite{REN200550, Sun2013, 10.1063/1.3072799, 10.1063/1.5119209, PhysRevB.86.064433, BURKOV2019540, PhysRevB.82.155124}. Note that the discrepancies among experimental results are attributed to defect scattering from sample-dependent Si vacancies \cite{PhysRevB.86.064433, PhysRevB.82.155124}, and our prediction serves as a theoretical lower bound for a defect-free crystal. Figure~\ref{Fig2}(b) presents the calculated $\kappa_{\rm L}$, $\kappa_{\rm e}$, and $\kappa_{\rm tot}$. The results are in reasonable agreement with experiments \cite{Sun2013}, with the 7\%$\sim$16\% overestimation of $\kappa_{\rm tot}$ arising from the existence of extrinsic scattering in experimental samples. This agreement validates our first-principles approach for separating $\kappa_{\rm e}$ and $\kappa_{\rm L}$. Notably, $\kappa_{\rm L}$ is appreciable over the entire $T$ range and dominates thermal transport at and below $T_{\rm room}$, accounting for 51\% of $\kappa_{\rm tot}$ (10 W/mK) at $T_{\rm room}$. This finding is anomalous against the conventional view of a negligible lattice contribution in metallic systems, echoing recent reports of large $\kappa$ in several metals \cite{PhysRevB.99.020305, KUNDU2020100214, chen2024origin, ding2024anharmonicity, kundu2021ultrahigh}.

With $\kappa_{\rm e}$ and $\kappa_{\rm L}$ determined, we then derive the electronic Lorenz number ($L_{\rm e}$). As clearly seen in Fig.~\ref{Fig2}(c), $L_{\rm e}$ consistently exceeds $L_{\rm 0}$ across the entire $T$ range, with the deviation reaching a maximum of $\sim$40\% at $T_{\rm room}$. This upward deviation signifies that charge carriers in CoSi transport thermal energy more effectively than electrical current. Conversely, if one assumes $\kappa_{\rm L}$to be negligible per conventional wisdom, the $L$ extracted from $\kappa_{\rm tot}$ deviates from $L_{\rm 0}$ by more than a factor of three (inset, Fig.~\ref{Fig2}(c)). It is thus evident that the common practice of applying the standard WF law to estimate $\kappa_{\rm tot}$ from $\sigma$ can be highly misleading in semimetals like CoSi. 

In weakly correlated electron systems, violations of the WF law typically manifest as a reduced Lorenz number ($L<L_{\rm 0}$), arising from the distinct effects of inelastic el-ph scattering processes on thermal and electrical transport \cite{PhysRevB.102.174306,pnas.2318159121,PhysRevB.100.144306}. This discrepancy occurs because electrons driven by an external electric field primarily undergo large-angle scattering to restore equilibrium, whereas those driven by a temperature gradient are effectively scattered by both large-angle and small-angle processes \cite{PhysRevB.102.174306}. In other words, electron scattering by large-wave-vector phonons is largely suppressed, making small-wave-vector phonons the dominant scattering source in charge transport. In contrast, both small- and large-wave-vector phonons can contribute to the scattering in electron energy transport via inelastic scattering. As a result, the momentum relaxation time tends to exceed the energy relaxation time. However, it is evident from Fig.~\ref{Fig2}(c) that under the constant relaxation time approximation, which assumes that the carrier lifetime is identical under electrical field and temperature gradient, the $L$ deviates more significantly upward from $L_{\rm 0}$, with a maximum deviation approaching 100\%. This suggests that although inelastic scattering tends to suppress $L$, a substantial upward deviation persists, thereby ruling out scattering-related mechanisms and pointing to the electronic band structure as its origin.

\begin{figure}
\includegraphics[scale=0.44]{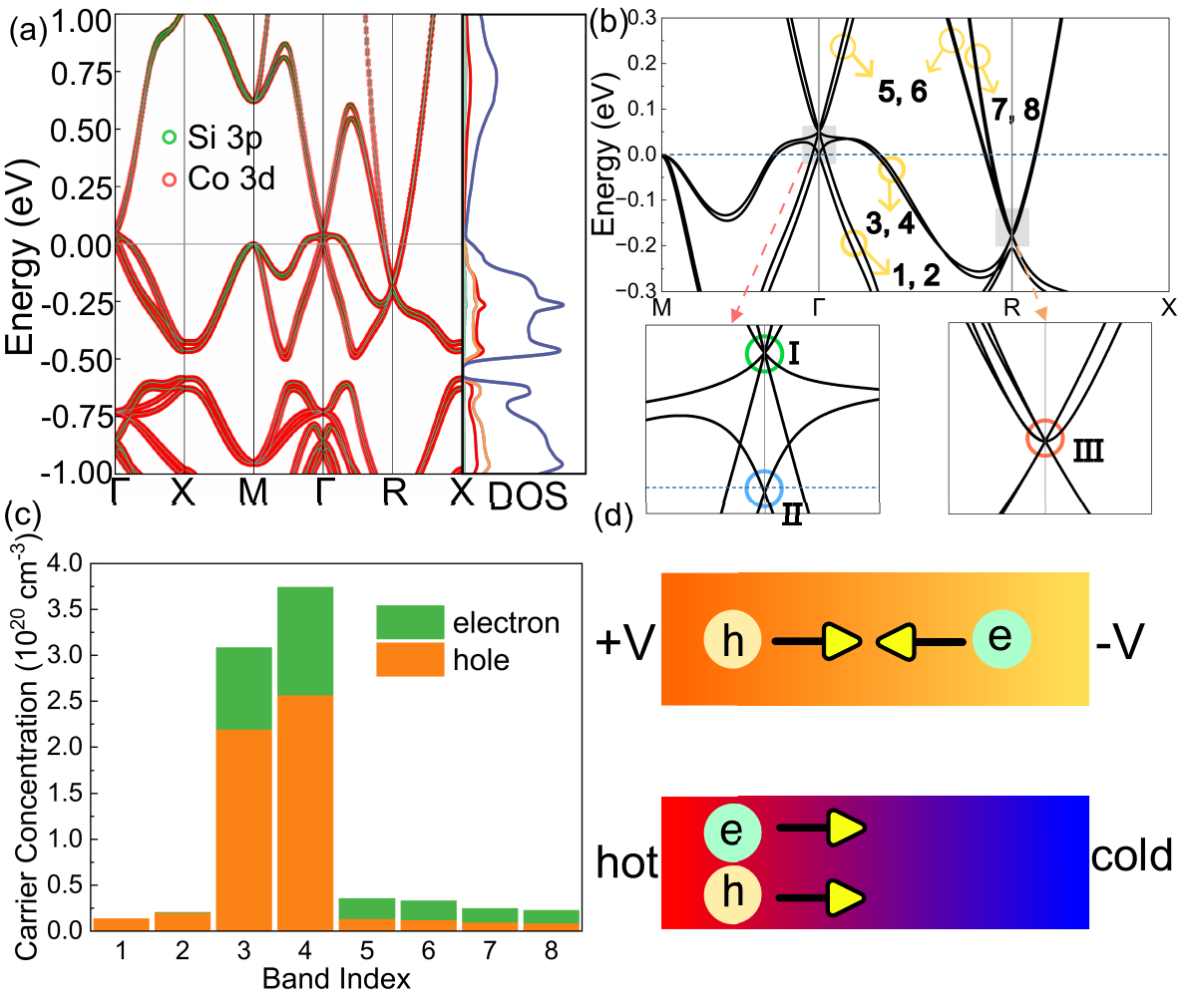}
\caption{(a) Electronic band structure and Orbital-projected density of states (DOS) of CoSi. The DOS decomposition shows contributions from different atomic orbitals: 
Si $p$ orbitals (green), Co $d_{z^2}$ and $d_{x^2-y^2}$ orbitals (orange), 
Co $d_{xy}$, $d_{yz}$, and $d_{xz}$ orbitals (red), and total DOS (blue).
The energy zero is set to the Fermi level. (b) Zoomed-in band structure near the Fermi level. (c) Band-resolved carrier concentration at 300 K. (d) Schematic illustration of carrier response to an electric field and a temperature gradient.}
\label{Fig3}
\end{figure}

Figure~\ref{Fig3}(a) presents the calculated electronic band structure and the corresponding projected DOS with SOC included. Near the Fermi level, the electronic states are dominated by the Co 3$d$ orbitals. The low electronic DOS is observed at the Fermi level, which accounts for the weak ph-el scattering, as its intensity is known to be positively correlated with the electronic DOS at the Fermi energy \cite{ding2024anharmonicity, kundu2021ultrahigh}. Given that electronic transport is governed by the band structure near the Fermi level, Fig.~\ref{Fig3}(b) shows a magnified view of this region, where the electronic states around the Fermi level—comprising hole-like pockets from bands 3 and 4 at $\bf \Gamma$ point and electron-like pockets from bands 5 to 8 at $\bf R$ point--—identify it as a compensated semimetal. This characteristic is a direct consequence of the symmetry-protected topological band structure. The $P2_13$ space group, with its nonsymmorphic screw symmetries, enforces threefold degeneracies at both the $\bf \Gamma$ and $\bf R$ points in the absence of SOC. Upon including SOC, these degeneracies are lifted into distinct topological nodes: at $\bf \Gamma$, into a fourfold degenerate fermion ($\mathrm{I}$) enforced by time-reversal symmetry and a twofold Weyl fermion ($\mathrm{II}$); at the time-reversal invariant $\bf R$ point, into a robust sixfold degenerate fermion ($\mathrm{III}$) protected by the interplay of nonsymmorphic and time-reversal symmetries \cite{PhysRevLett.119.206402, doi:10.1126/science.aaf5037}. The Fermi energy is pinned such that it lies just below the fourfold node at $\bf\Gamma$, depleting electrons from the valence band to form hole pockets, and just above the sixfold node at $\bf R$, doping electrons into the conduction band to form electron pockets, thereby realizing the compensated semimetallic state. These symmetry-protected band characteristics collectively yield comparable electron and hole concentrations across individual bands (bands 3–8), as illustrated in Fig.~\ref{Fig3}(c). The emergence of such approximate electron–hole symmetry enables bipolar transport. The corresponding physical picture is sketched in Fig.~\ref{Fig3}(d): under an external electric field, electrons and holes drift in opposite directions, whereas under a temperature gradient, both carriers flow together from the hot to the cold side. Consequently, the spontaneous flow of electrons and holes generates an additional energy current while maintaining zero net electric current. This extra energy current arises from the energy extracted from the hot side during electron-hole separation and the energy released to the cold side upon recombination.

\begin{figure}
\includegraphics[scale=0.42]{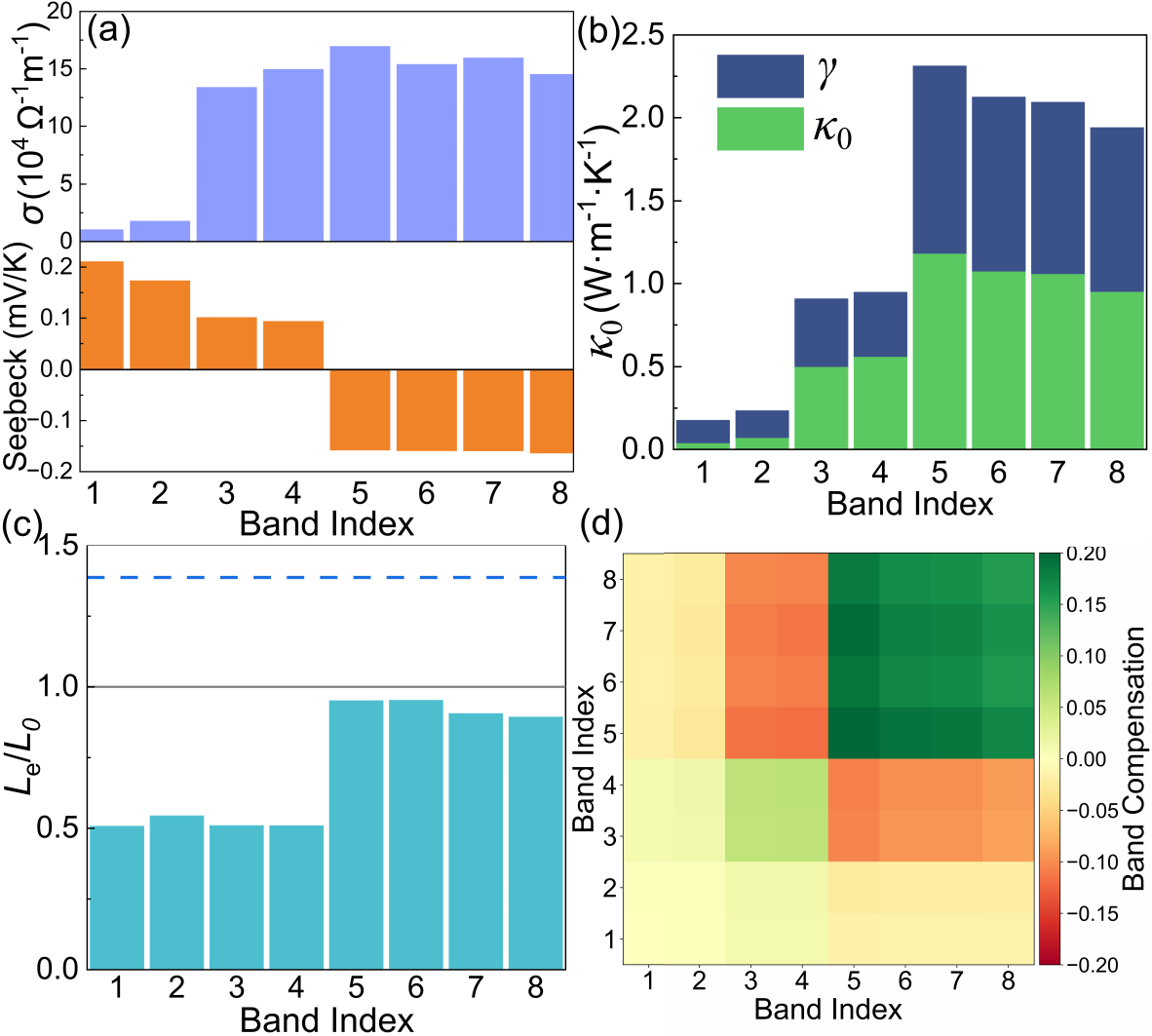}
\caption{The contribution of each band to (a) the electrical conductivity and Seebeck coefficient, (b) the electronic thermal conductivity, and (c) the Lorenz number at $T_{\rm room}$. The blue dashed line in (c) denotes the Lorenz number in the multi-bands case. (d) The carrier thermal conductivity term arising from the Seebeck effect at 300 K.}
\label{Fig4}
\end{figure}

To characterize the contribution of bipolar transport, we further analyze the band-resolved transport properties near the Fermi level within the electron BTE \cite{PhysRevB.94.085204}. Assuming a single-band scenario where only the $n$-th band contributes to the electronic transport, the $\kappa_{\rm e}$ at zero electric current for a cubic system is given by
\begin{eqnarray}
(\kappa_e)_n =&&  (\kappa_0)_n - \frac{\alpha_n (\sigma S)_n}{\sigma_n} 
\label{eq1}
\end{eqnarray}
where $\sigma_n$, $S_n$ denote the electrical conductivity and Seebeck coefficient, and $\alpha_n$ denote the product of electrical conductivity and Peltier coefficient, with their full expressions given in the Supplemental Material \cite{suppl}. The first term ($\kappa_{\rm 0}$) in Eq.~\ref{eq1} represents the short-circuit $\kappa_{\rm e}$, and the second term (denoted as $\gamma$) accounts for the suppression of the Seebeck effect in the open-circuit condition \cite{grimenes2025commonerrorsboltztrapbasedcalculations}. Band-resolved analysis in Fig.~\ref{Fig4}(a) reveals that bands 3--8 contribute the majority of the electrical conductivity, and the Seebeck coefficient is characterized by hole-like contributions from bands 1--4 and electron-like contributions from bands 5--8. The decomposition of $\kappa_{\rm e}$, $\kappa_{\rm 0}$, and $\gamma$ across bands are displayed in Fig.~\ref{Fig4}(b), showing that $\kappa_{\rm e}$ and $\gamma$ are comparable in magnitude for all bands, revealing a strong thermoelectric suppression of thermal conductivity. Consequently, the resulting Lorenz number for each band falls consistently below $L_0$ and the Lorenz number of multi-bands, as shown in Fig.~\ref{Fig4}(c). This finding clearly attributes the anomalously large Lorenz number observed in Fig.~\ref{Fig2}(c) to multi-bands effects.

Different from the single-band scenario, the electronic thermal conductivity in multi-bands case is given by
\begin{eqnarray}
\kappa_e =&&  \sum_n (\kappa_0)_n - \frac{\sum_m \alpha_m \sum_n (\sigma S)_n}{\sum_n \sigma_n} 
\label{eq2}
\end{eqnarray}
which gives rise to interband cross-terms. When both types of charge carriers coexist across different bands, these cross-terms can constructively enhance $\kappa_{\rm e}$. For instance, in a two-band model, the scalar expansion of this term takes the form: $\gamma =\frac{\sum_{mn}\alpha_m (\sigma S)_n}{\sigma_{\rm tot}} = \frac{\alpha_1 (\sigma S)_1}{\sigma_1+\sigma_2} + \frac{\alpha_2 (\sigma S)_1}{\sigma_1+\sigma_2} + \frac{\alpha_1 (\sigma S)_2}{\sigma_1+\sigma_2} + \frac{\alpha_2 (\sigma S)_2}{\sigma_1+\sigma_2}$. Here, the second and third terms contribute positively to $\kappa_{\rm e}$ whenever the two bands host opposite types of charge carriers. In our system, the non-additive behavior in the multi-band transport is even more pronounced. Figure~\ref{Fig4}(d) shows the band compensation matrix ($\gamma_{mn}= \frac{\alpha_m (\sigma S)_n}{\sigma_{\rm tot}}$), revealing that the strong enhancement in $\kappa_e$ originates from negative compensation elements between bands 3, 4 and bands 5--8, which can be attributed to electron-hole compensation between the corresponding hole and electron pockets (Fig.~\ref{Fig3}(b))---whose Seebeck coefficients have opposite signs (Fig.~\ref{Fig4}(a)) along with the substantial conductivity contributions from bands 3–-8. This finding clearly demonstrates that the anomalously large Lorenz number in CoSi arises from these interband cross-terms, which originate from a unique band structure dictated by its topological nature. We propose that this topology-induced bipolar enhancement of thermal conductivity should be universal across a wide range of topological semimetals with strong electron-hole compensation.

In summary, we have employed first-principles Boltzmann transport calculations to probe the electrical and thermal conductivities of the topological semimetal CoSi. Our results reveal an anomalously large $\kappa_{\rm L}$, which contributes more than 50\% to the total thermal conductivity at $T_{\rm room}$, contrary to the established belief that $\kappa_{\rm L}$ is negligible in metals. More importantly, we observe a significant upward violation of the WF law, with the electronic Lorenz number ($L_{\rm e}$) deviating from $L_0$ by up to ~40\%. Detailed band-resolved analysis reveals that this departure stems from electron-hole compensation driven by a topologically dictated band structure, which gives rise to pronounced bipolar diffusive transport. Our findings highlight the  topological band-induced anomalous thermal transport behavior, which is expected to be ubiquitous in other topological semimetals with comparable electron-hole Fermi pockets.  

\textit{Acknowledgements.} This work is supported by the National Natural Science Foundation of China (Grant No. 12374038 and No. 12404045), Fundamental Research Funds for the Central Universities of China (Grant No. 2023CDJKYJH104), and Chongqing Natural Science Foundation (Grant No. CSTB2022NSCQ-MSX0834). 

\nocite{*}
\bibliographystyle{apsrev4-2}
\bibliography{main}

\end{document}